# Observation of off-diagonal geometric phase in polarized neutron interferometer experiments

Y. Hasegawa[1], R. Loidl[2], G. Badurek[1], M. Baron[2], N. Manini[3], F. Pistolesi[4*] and H. Rauch[1]

[1]Atominstitut der Österreichischen Universitäten,
Stadionallee 2, A-1020 Wien, Austria

[2]Institut Laue Langevin, B. P. 156, F-38042 Grenoble Cedex 9, France

[3]Dipartmento di Fisica, Università di Milano, Via Celoria 16, I-20133 Milano, Italy and INFM, Unità di Milano, Milano, Italy

[4]European Synchrotron Radiation Facility, B.P. 220, F-38043, Grenoble, France

## Abstract

Off-diagonal geometric phases acquired in the evolution of a spin-1/2 system have been investigated by means of a polarized neutron interferometer. Final counts with and without polarization analysis enable us to observe simultaneously the off-diagonal and diagonal geometric phases in two detectors. We have quantitatively measured the off-diagonal geometric phase for noncyclic evolutions, confirming the theoretical predictions. We discuss the significance of our experiment in terms of geometric phases (both diagonal and off-diagonal) and in terms of the quantum erasing phenomenon.

---

[*] Present address:
Laboratoire de Physique et Modélisation des Milieux Condensés,
Centre National de la Recherche Scientifique,
BP 166, F-38042 Grenoble Cedex 9, France

# I. INTRODUCTION

Consider the dynamics of a quantum system governed by a parameterized Hamiltonian. When the parameter is varied along a curve, a normalized eigenstate, with energy eigenvalue $E(\mathbf{s})$, acquires, in addition to the standard dynamical phase given by $\exp\left[-(^i/_\hbar)\int E(\mathbf{s})d\mathbf{s}\right]$, a phase factor. This extra phase is called "geometric", since it is independent of the Hamiltonian, of the energy of the eigenstate and of the rate of the evolution, and it only depends on the sequence of quantum states. This geometric phase was observed by Pancharatnam in the interference experiment of polarized light in the fifties [1]. Berry then formulated the geometric phase for adiabatic transport along a closed curve (i.e. cyclic evolution [2]), attracting considerable interest [3]. The concept of the geometric phase introduced by Berry was generalized, for instance, to nonadiabatic [4] and noncyclic [5] evolutions. Many experiments in a wide range of physical phenomena were reported to determine the geometric phase [3]: in particular, the geometric phase for the most basic and fundamental example, the two-level system constituted by the neutron spin 1/2, was measured in early neutron experiments [6,7]. The geometric phase can play an important role in quantum information processing: a controlled phase gate, which requires conditional quantum dynamics, was demonstrated in a nuclear magnetic resonance experiment with the use of a conditional geometric phase [8,9].

The Berry-Pancharatnam geometric phase is the phase of the scalar product of the parallel-transported initial and final states $\langle\Psi^{//}(\mathbf{s}_1)|\Psi^{//}(\mathbf{s}_2)\rangle$ [2,10]. It may happen that the parallel-transported state is orthogonal to the initial state: in this case the geometric phase is undefined. Recently, the concept of the geometric phase has been extended to the phase factors coming from *cross* scalar products $\langle\Psi_j(\mathbf{s}_1)|\Psi_k(\mathbf{s}_2)\rangle$ of nondegenerate eigenstates $|\Psi_i\rangle$ [11]. The phase of suitable combinations of these cross products turn out to be a well-defined geometric object, even when the conventional "diagonal" geometric phase is undefined. With respect to diagonal products, these cross products carry complementary geometric information on the evolution and, accordingly, such geometric phases have been termed

"off-diagonal." This formalism was applied successfully to a deformed microwave resonator experiment [12,13].

Neutron optical experiments, based on interference of matter waves, have provided elegant demonstrations of the foundations of quantum mechanics [14,15]. In particular, neutron interferometry experiments have made an impact for more than two decades [16-19]. Among them, those with a polarized incident beam served as an ideal tool to investigate properties of a spin-1/2 system [20]. For instance, starting from spinor superposition [21,22], double resonance flipper experiments [23], geometric and dynamical phase measurements [24,25], and noncyclic Pancharatnam phase measurement [26] were accomplished. Besides the interferometric method, the geometric and dynamical phases are more precisely measured with the use of a neutron polarimeter [27]. A neutron interferometer with coupled interference loops was used to create and to measure the spin-independent geometric phase in a sort of double-slit experiment [28].

In this paper, we describe a neutron interferometric experiment that clearly exhibits and quantitatively measures the off-diagonal geometric phase acquired by noncyclic spinor evolutions of a spin-1/2 system. An incident polarized neutron beam splits into two beam paths and each spinor is rotated to induce appropriate spinor evolutions. With the help of a polarization analysis, the noncyclic off-diagonal geometric phases emerges straightforwardly as the shift of an intensity modulation, whereas the measurement without polarization analysis enables the observation of the diagonal geometric phase, i.e. the noncyclic Pancharatnam phase. Our data confirms theoretical predictions on the off-diagonal geometric phase illustrating its role for a generic evolution and its special significance when the diagonal geometric phase is undefined. In addition, we discuss the quantum-eraser phenomenon for these experiments. A brief report of part of the experimental results has been published previously [29].

## II. PRINCIPLE OF THE EXPERIMENT

### A. Off-diagonal geometric phase of spin-1/2 system

Consider two nonorthogonal quantum states $|\Psi_I\rangle$ and $|\Psi_{II}\rangle$. The standard interferometric way to measure the phase $\gamma$ of the scalar product $\langle\Psi_I|\Psi_{II}\rangle = \exp(i\gamma) |\langle\Psi_I|\Psi_{II}\rangle|$ is to superpose the two states inserting a controlled extra phase factor $\exp(i\chi)$:

$$|\Psi\rangle = \exp(i\chi) |\Psi_I\rangle + |\Psi_{II}\rangle \qquad (1)$$

The resulting interfering intensity

$$I = \langle\Psi|\Psi\rangle = \langle\Psi_I|\Psi_I\rangle + \langle\Psi_{II}|\Psi_{II}\rangle + 2 |\langle\Psi_I|\Psi_{II}\rangle| \cos(\chi-\gamma) \qquad (2)$$

oscillates as a function of the phase angle $\chi$. The offset of the cosine oscillation of I measures $\gamma$, and the amplitude of these oscillations measures the overlap $|\langle\Psi_I|\Psi_{II}\rangle|$.

The basic interferometric concept sketched here can be applied to compare states evolving in time according to a given dynamics. The most basic quantum system is an S=1/2 spinor. In this paper we use neutron interferometry to measure the phases acquired by the neutron spin state precessing under the action of suitably arranged magnetic fields and polarizers. A beam of neutrons initially in a pure $|\Psi^+\rangle = \begin{bmatrix} \cos(\theta/2) \\ \sin(\theta/2) \end{bmatrix}$ polarization state is split and, along the two paths, undergoes different transformations represented by two linear operators A and B (represented by 2 by 2 unitary matrices), before the two phase-coherent beams $|\Psi_I\rangle = A |\Psi^+\rangle$ and $|\Psi_{II}\rangle = B |\Psi^+\rangle$ are brought together at the final counter.

By choosing suitably A and B in the setup it is possible to measure the recently-proposed off-diagonal geometric phase [11] for the evolution of two states according to a given unitary evolution operator U. The off-diagonal phase factor is defined by

$$\exp(i\, \gamma^{\text{off}}_{+-}) = \Phi(\langle\Psi^+|U|\Psi^-\rangle \langle\Psi^-|U|\Psi^+\rangle) , \qquad (3)$$

where $\Phi(z)=z/|z|$, $|\Psi^-\rangle = \begin{bmatrix} -\sin(\theta/2) \\ \cos(\theta/2) \end{bmatrix}$ is the spinor orthogonal to $|\psi^+\rangle$ and U is the unitary operator representing the evolution of the states *according to parallel transport*. Specifically, $U=U(s_2)$ governs the evolution $|\Psi^\pm(s)\rangle$ of $|\Psi^\pm\rangle$ to the final value $s_2$ of the evolution parameter s, according to

$$|\Psi^{\pm}(s)\rangle = U(s) |\Psi^{\pm}\rangle, \qquad (4)$$

with the parallel transport condition on the phases along the evolution such that the following equation holds:

$$\langle \Psi^{\pm}(s)|\nabla_s|\Psi^{\pm}(s)\rangle = 0. \qquad (5)$$

In this way, the (non geometric) dynamical phase vanishes identically along the evolution. The choice of A and B that permits to measure $\gamma_{+-}^{off}$ interferometrically is the following:

$$A = P(|\Psi^-\rangle) U^{-1}, \qquad B = P(|\Psi^-\rangle) U, \qquad (6)$$

where $P(|\Psi^-\rangle) = |\Psi^-\rangle\langle\Psi^-|$ is a projection operator (representing a neutron spin polarizer), and $U^{-1}$ is the reversed evolution with respect to U.

In the present work, we only consider precessions around the $\hat{z}$ direction, determining a unitary time evolution

$$U(t) = \begin{bmatrix} \exp(i\omega_L t/2) & 0 \\ 0 & \exp(-i\omega_L t/2) \end{bmatrix}, \qquad (7)$$

where $\omega_L$ is the Larmor frequency: with this special choice, spinors restricted to the $\hat{x}$–$\hat{y}$ plane ($\theta=\pi/2$) are parallel-transported by U. In addition, even for ($\theta \neq \pi/2$), the extra dynamic phases $\Phi_D^+ = -\Phi_D^- = \frac{\omega_L t}{2} \cdot \cos\theta$ introduced by U happen to cancel exactly in the product (3), and are therefore immaterial for $\gamma_{+-}^{off}$.

Under the action of such choice of evolution operators, the expected intensity

$$I = \||\exp(i\chi) |\Psi_I\rangle + |\Psi_{II}\rangle\|^2 \qquad (8)$$

can be computed in accordance to (2). Applying explicitly the unitary evolution U of Eq. (7), the interfering states $|\Psi_I\rangle = P(|\Psi^-\rangle) U^{-1} |\Psi^+\rangle = i \sin(\theta) \sin(\omega_L t/2) |\Psi^-\rangle$ and $|\Psi_{II}\rangle = P(|\Psi^-\rangle) U |\Psi^+\rangle = -|\Psi_I\rangle$ combine to give

$$I = \langle\Psi^+|U|\Psi^-\rangle\langle\Psi^-|U^{-1}|\Psi^+\rangle + \langle\Psi^+|U^{-1}|\Psi^-\rangle\langle\Psi^-|U|\Psi^+\rangle +$$

$$+ 2 \, \text{Re}[\exp(i\chi) \langle\Psi^+|U|\Psi^-\rangle\langle\Psi^-|U|\Psi^+\rangle]$$

$$= 2 \sin^2(\theta) \sin^2(\omega_L t/2) - 2 \sin^2(\theta) \sin^2(\omega_L t/2) \cos(\chi)$$

$$= 2 \sin^2(\theta) \sin^2(\omega_L t/2) [1 + \cos(\chi-\pi)] . \qquad (9)$$

This derivation illustrates that the inverse evolution $U^{-1}$ fulfills the role of eliminating all U-dependence (apart from a global factor $\sin^2(\omega_L t/2)$) in the non-interfering amplitude, while letting the phases of the off-diagonal matrix elements of U combine in the correct way to have the off-diagonal geometric phase $\gamma_{+-}^{\text{off}}$ of Eq. (3) as a shift in the interfering amplitude.

The comparison of Eqs. (9) and (2) indicates that $\gamma_{+-}^{\text{off}} = \pi$, an equation representing the phase relation $|\Psi_{II}\rangle = -|\Psi_I\rangle$. We have recovered here the result of Ref. [11], where it was shown that for *any* evolution (not necessarily restricted to precessions around the $\hat{\mathbf{z}}$ direction) the off-diagonal phase $\gamma_{+-}^{\text{off}}$ of the two-state system must equal $\pi$. This is a consequence of the relatively restrict dynamics of the two-state system. Systems with a larger number of degrees of freedom are expected to show a more intricate pattern of diagonal and off-diagonal phases, depending explicitly on the states considered and on the evolution path.

**B. Experimental strategy with the use of neutron interferometer**

Equation (8) describes mathematically the following experiment: a neutron beam, fully polarized to the incident spinor $|\Psi^+\rangle$, is split (the 2-terms sum), then appropriate spinor rotations U and $U^{-1}$, in addition to the auxiliary phase shift of $\chi$, act on the two paths. The two beams are brought together and finally the intensity is measured with the polarization analysis orthogonal to the incident spinor (the projector factor). The unitary evolution is realized by the magnetic field in the $+\hat{\mathbf{z}}$ direction. The inverse precession $U^{-1}$ is simply obtained by applying a reversed field to one of the two arms of the interferometer.

We draw the reader's attention to the fact that, when the spin analysis is not performed (A=U, B=$U^{-1}$), Eq. (8) reduces to

$$I = \| [\exp(i\chi) U^{-1} + U] |\Psi^+\rangle \|^2 = \| [\exp(i\chi) + U(2t)] |\Psi^+\rangle \|^2 \qquad (10)$$

This exhibits another possibility to observe the diagonal part of the geometric phase, i.e. the noncyclic Pancharatnam phase, simply by bypassing the final spin analysis in the experiment. Note however the presence of $U(2t)=UU=U^2$ in the Eq. (10), indicating that the measured phase refers to a path undergoing *twice* the precession angle experienced by the actual neutron spin of either split beam.

### C. Poincaré sphere description

A completely general geometric interpretation of the off-diagonal geometric phase was given in Ref. [11] in terms of geodesic (shortest path connecting two points, associated to null geometric phase) in the projective space. The off-diagonal geometric phase $\gamma_{jk}^{\Gamma}$ can be represented as an integral over the surface bounded by the four paths: $\Gamma_j$ (trajectory of the state $|\Psi_j\rangle$) + $G_{jk}$ (geodesic path from $|\Psi_j(s_2)\rangle$ to $|\Psi_k(s_1)\rangle$) + $\Gamma_k$ (trajectory of the state $|\Psi_k\rangle$) + $G_{kj}$ (geodesic from $|\Psi_k(s_2)\rangle$ to $|\Psi_j(s_1)\rangle$); similarly, the diagonal geometric phase, for instance $\gamma_j^{\Gamma}$, is given by a surface enclosed by the two paths $\Gamma_j$ + $G_{jj}$ (geodesic from $|\Psi_j(s_2)\rangle$ to $|\Psi_j(s_1)\rangle$).

In the special case at hand of a two-state system, the projective space is conveniently represented by the Poincaré (unit) sphere, for two main reasons. First, this mapping carries over the metric so that any projective geodesic is mapped on a spherical geodesic, i.e. the shortest path connecting two points. Second, the integral [2] over the surface bounded by a closed path in projective space that gives the geometrical phase acquired along such path, reduces simply to twice the geometric surface enclosed by the loop on the Poincaré sphere.

Initially, we take the incident spinor $|\Psi^+\rangle$ pointing in a direction tilted by $\theta$ from $+\hat{\mathbf{z}}$ in the $\hat{\mathbf{z}}$–$\hat{\mathbf{y}}$ plane, then let a magnetic field parallel to $\hat{\mathbf{z}}$ rotate it by an angle $\alpha=\omega_L t$. The orthogonal spinor $|\Psi^-\rangle$ spin points the opposite way, at each time along the evolution. In Fig.1 we represent the initial spinor $|\Psi_i^+\rangle=|\Psi^+\rangle$ (bold solid arrow) evolving to $|\Psi_f^+\rangle$ (bold dashed arrow) under the action of the magnetic field, thus defining the trajectory $\Gamma_+$. The diametrically opposed evolution of $|\Psi^-(t)\rangle$ from $|\Psi_i^-\rangle$ to $|\Psi_f^-\rangle$ gives the trajectory $\Gamma_-$. For any $\alpha$, the initial and final states lie in a unique plane, except in the special case $\alpha=0°$ mod $360°$, for which $\gamma_{+-}^{\Gamma}$ is undefined, since the off-diagonal elements of U vanish. For any other value of α, the geodesics $G_{+-}$ and $G_{-+}$ indicated in Fig. 1 are uniquely determined.

The shaded area surrounded by $\Gamma_{+}$, $G_{+-}$, $\Gamma_{-}$ and $G_{-+}$, equals one half of the off-diagonal geometric phase. By recognizing the equivalence of the shaded area with the half sphere surrounded by the sequence of the four geodesic paths $G_{++}$, $G_{+-}$, $G_{--}$, and $G_{-+}$, it is clear that this shaded area equals half the spherical surface, i.e. 2π (independently of α). This illustrates the constant value $\gamma_{+-}^{\text{off}}=\pi$ taken by the off-diagonal geometric phase, as discussed in Sect. A. The Poincaré sphere representation also shows how the off-diagonal geometric phase is especially significant when the diagonal geometric phase is ill defined. The diagonal Berry-Pancharatnam phase $\gamma_{+}$ is represented in Fig. 1 by the spherical sector between $\Gamma_{+}$ and $G_{++}$. This sector becomes ill defined for θ=90° and α=180°, since $G_{++}$ is then arbitrary. Nevertheless, the off-diagonal geometric phase is deduced independently of both geodesics $G_{++}$ and $G_{--}$, therefore its value remains well defined. Incidentally, in this special case the geodesic paths $G_{+-}$ and $G_{-+}$ reduce to single points.

### III. EXPERIMENTAL

### A. Setup

The experiment was carried out at the perfect crystal interferometer beam line S18 at the high flux reactor at the Institut Laue Langevin (ILL) [30]. A schematic view of the experimental setup is shown in Fig.2. Before falling on a skew-symmetric triple-Laue interferometer, the incident neutron beam was collimated and monochromatized by the 220 Bragg reflection of a Si perfect crystal monochromator placed in the thermal neutron guide H25. The wave length was tuned to give a mean value $\lambda_0$=1.92Å ($\Delta\lambda/\lambda_0$~0.02). This incident beam was slightly separated in angle with the use of the spin-dependent birefringence of neutrons on passing through prismatically shaped magnetic fields oriented in the vertical direction, i.e. perpendicular to the beam trajectory defining $\hat{z}$ [31]. The beam cross section was confined to 8×8 mm$^2$ and a spin rotator F was inserted to tune the polar angle θ of the incident spinor. The interferometer was adjusted to give 220 reflection in a non-dispersive arrangement relative to the monochromator, it picking up neutrons polarized in one direction. A pair of water-cooled Helmholtz coils produced a fairly uniform magnetic guide field $B_0\hat{z}$ across the interferometer. An isothermal box enclosed the interferometer to achieve reasonable thermal environmental isolation. A magnetically saturated Heusler crystal together with a Larmor accelerator and a rectangular spin-1/2 turner enabled spin analysis, i.e. the selection of neutrons with certain polarization directions, for one of the two interfering beams [32]. This O-beam was used to observe the phase shift due to the off-diagonal geometric phase. In addition, we measured the intensity of the other interfering (H-)beam without spin analysis to show the phase shift due to the diagonal geometric phase.

In order to estimate depolarization, the efficiency of the spinor rotations in the spin analyzing part, and other imperfections of the setup, the interferometer was first adjusted to accept only spin-up polarized neutrons, i.e. $|+z\rangle$, with the spin rotator F turned off. The fraction of the $|+z\rangle$ component was measured to be 0.92 with the use of the spin turner. Since the present measurement demands the polarization of the incident beam to lie in the $\hat{x}$–$\hat{z}$ plane, the spin rotator F was turned on to tune the polar angle θ, of the spin direction. The magnetic guide field was optimized to $B_0$=18 G to avoid additional depolarization. The measurement with the use of the Larmor accelerator revealed that the fraction of the desired component was reduced to 0.87, which was still high enough to accomplish the measurements. Nonessential spinor precessions are induced by the guide field all through the beam trajectories. Nevertheless, the commutability between the operators for the essential spinor rotations (U and $U^{-1}$ in Eq.(8)) and the nonessential rotations (introduced by the guide field) allows to compensate for these nonessential rotations behind the interferometer, with the use of the Larmor accelerator.

The essential unitary evolutions, U and $U^{-1}$, were realized in the experiment by setting the magnetic fields in the two identical spin rotators in an anti-parallel direction, i.e. in the $\pm\hat{z}$ directions. In practice, a pair of spin rotators I and II was inserted in the split beam paths. These spin rotators were identical DC-coils except for the current directions, i.e. the directions of the induced magnetic fields. They were connected in series to give the same strengths of the magnetic fields in the $\pm\hat{\mathbf{z}}$ direction. Each coil was made of a 8 mm wide and 0.5 mm thick anodized Al band, where the hydrogen in the insulator was replaced by deuterium to avoid small angle scattering, wound onto a water-cooled Cu frame. These coils induced 180° spinor rotations when operated at 9.8 A dissipating about 0.6 W each. The rotation angle $\alpha=\omega_L t$ was tuned by adjusting the current and possibly by changing their polarities. Special attention was paid to eliminate both heat and vibration transmission to the interferometer.

Interferograms were obtained in two detectors by rotating a 5 mm thick parallel-sided Al plate as a phase shifter. The detectors were placed downstream of the interfering O-beam in the forward direction behind a spin analysis system, and directly in the H-beam in the reflected direction without a spin analysis. These two detectors enabled us to measure simultaneously the properties of the diagonal and off-diagonal geometric phases. For each phase-shifter position, two intensities were measured: typically with the spin rotators I and II turned on (measuring time 10 second), and with the spin rotators turned off (measuring time 20 second). This procedure gave simultaneously two interferograms with and without spinor rotations, thus avoiding instabilities producing undesired phase shifts between the two measurements.

In measuring the interferograms, we determined the phase shift, given by the phase shifter, to have the same value at the same position, when the coils are not activated. In practice, the imperfect polarization of the beam enabled to record interferograms even for the unaffected beam, i.e. without spinor rotations, where no modulation is expected from the theory. The contrast of the interference oscillations, obtained by the O-beam, was typically 64% for the empty interferometer, and was reduced to about 40% after inserting the dual spin rotators. This contrast dropped in some situations to about 30%, mainly due to thermal disturbances. We repeated the same measurements at least three times to ensure that the results were reliable.

### B. Interferograms

In the first experiment, the polar angle of the incident polarization was tuned to $\theta=90°$, which shows the most characteristic property of the off-diagonal geometric phase. The spin rotators I and II were tuned to give rotations, $\alpha=45°$, $\pm67.5°$, $\pm90°$, $\pm180°$, and $\pm225°$. The Larmor accelerator was adjusted, so that the 180° precession was induced to analyze the spinor orthogonal to the incident one. In addition, the nonessential precession by the guide field was compensated. Figure 3 shows typical interferograms for several spinor rotation angles $\alpha$, together with the original interferogram ($\alpha=0°$) as a reference. The curves are least squares fits.

In the interferograms of the O-beam (Fig.3(a)), where the off-diagonal geometric phase is expected to appear, all patterns with spinor rotations are reflected. This confirms the phase shift of 180° due to the off-diagonal geometric phase. In addition, one sees that the amplitude of the oscillations has its maximum at $\alpha=180°$ and that it falls gradually when $\alpha$ departs from 180°. This arises from the fact that, with the increase of the spinor rotation angle, the orthogonal component to the incident spinor increases until $\alpha=180°$ and then decreases. In contrast, interferograms of the H-beam (Fig.3(b)) show phase shifts of 0° or 180°. These shifts result from the Pancharatnam phase, or from a noncyclic *pure* diagonal geometric phase, since the dynamical phase, or the parallel-transport variant phases $\Phi_D^{\pm}$ is zero in these particular circumstances. It is worth noting here that no interference oscillations emerged, when the spinor rotation angle $\alpha=\pm90°$. In the vicinity of this point, a phase jump from 0° to 180° is expected to occur, which is characteristic for circuits around phase singularities [33]. We will discuss this no-oscillation in more detail with regard to the significance of the off-diagonal geometric phase as well as the quantum eraser in Sect. IV. In addition, one sees here the sign changes of the spinor wave function with $\pm360°$ spinor rotations (i.e. for $\alpha=\pm180°$), as resulting from the $4\pi$-periodicity of the spin-1/2 wave function.

In a second series of measurements, the angle characterizing the incoming polarization was varied, to $\theta=135°$, 60°, and 30°. At each polar angle, we induced spinor precessions $\alpha=\pm90°$, $\pm180°$, and $\pm225°$. The Larmor accelerator was adjusted to analyze the spinor orthogonal to the incident one. Typical interferograms are shown in Fig.4 for spinor rotation angle $\alpha=60°$ together with the original interferogram ($\alpha=0°$) as a reference. The curves are least squares fits. Also here, the O-beam (Fig.4(a)) shows the phase shift of 180° due to the off-diagonal geometric phase as the theory predicts, whereas the H-beam (Fig.4(b)) indicates a gradual shift of the oscillations. In this case, this shift is not due to a pure noncyclic diagonal geometric phase, but to the combination of the noncyclic dynamical and geometrical phases, or rather the noncyclic Pancharatnam phase. The sign changes of the spin-1/2 wave functions with $\pm 360°$ spinor rotations, i.e. $\alpha = \pm 180°$, are also seen.

### C. Evaluation of off-diagonal geometric phases

In the previous subsection, the variations of the interferograms qualitatively showed the effect of the off-diagonal geometric phase. Here, the noncyclic off-diagonal geometric phase is quantitatively derived from the experimental data. The shifts of the oscillations obtained by the O-beam is exactly the measure of the off-diagonal geometric phase, when one extracts it as the shift from the original interferogram ($\alpha = 0°$). The observed off-diagonal geometric phases at various spinor rotation angle, $\alpha$, are plotted for the cases $\theta = 135°$, $90°$, $60°$, and $30°$ in Fig.5. For $\theta = 30°$ and $\alpha = 90°$, the amplitude of the oscillation was too small to give a reliable value; thus these cases were discarded. The theory predicts a shift of $180°$ for all cases. The experimental results show good agreement with the theory. Slight deviations from the value of $180°$ seem systematic, and we supposed them to be due to the imperfect polarization. Theoretical calculations of the off-diagonal geometric phase as should be observed in the experiment, accounting for the effect of the actual polarization, are indicated as dashed lines in the Fig.5. The better agreement between predicted and observed results justify a posteriori these considerations and the experimental results.

## IV. DISCUSSION

### A. Off-diagonal matrix elements and phases

It is instructive to show the unitary evolution U with the $|\pm z\rangle$ spinor basis, which will clarify the significance of the quality measured in the experiment, i.e. the off-diagonal property. The transformation of the basis from the $|\pm z\rangle$ to the $|\Psi^\pm\rangle$ spinor basis is given in a matrix form by

$$T = \begin{bmatrix} \cos(\theta/2) & \sin(\theta/2) \\ -\sin(\theta/2) & \cos(\theta/2) \end{bmatrix}. \tag{11}$$

With this transformation matrix, the unitary evolution reads

$$\begin{aligned} U' &= TUT^{-1} \\ &= \begin{bmatrix} \cos(\omega t/2) + i\cos\theta\cdot\sin(\omega t/2) & -i\sin\theta\cdot\sin(\omega t/2) \\ -i\sin\theta\cdot\sin(\omega t/2) & \cos(\omega t/2) - i\cos\theta\cdot\sin(\omega t/2) \end{bmatrix}. \end{aligned} \tag{12}$$

Equation (3) yields $\exp[i\{\gamma_{+-}^{\text{off}} + (\Phi_D^- + \Phi_D^+)\}] = \Phi(U'_{-+} \cdot U'_{+-})$, and the present experiments measured exactly the value of $\Phi(U'_{-+} \cdot U'_{+-})$, which is deduced from the off-diagonal elements of the unitary evolution $U'$. We recall that in our setup the sum of the dynamical phases, or the parallel-transport variant phases, $\Phi_D^- + \Phi_D^+ = 0$. Therefore, the value $\Phi(U'_{-+} \cdot U'_{+-})$, representing the phase of the product of the off-diagonal elements, directly corresponds to the off-diagonal geometric phase $\gamma_{+-}^{\text{off}}$.

The off-diagonal geometric phase has its greatest importance, when the diagonal geometric phase is undefined, i.e. with the 180° spinor rotation in the case of the polar angle of the incident polarization $\theta = 90°$. In our setup, the H-beam without the spin analysis allowed to observe the diagonal element of the geometric phase, only for a doubled spinor rotation angle $2\alpha$. Figure 6 shows the intensity modulations in the measurements of the diagonal and off-diagonal geometric phases when the relevant spinor rotation angle is 180°. The modulations of the diagonal and the off-diagonal components were obtained by the H-beam with $\alpha = -90°$, and the O-beam with the $\alpha = -180°$, respectively. It is clearly seen that, even when the diagonal one is undefined due to the absence of the oscillation, the off-diagonal geometric phase is defined, and carries the geometrical information associated to the evolution of the state.

### B. Quantum erasing

As shown above, both off-diagonal and diagonal geometric phases were recorded with the use of the O-beam with the spin analysis and the H-beam without spin analysis in our experiment. In respect to the wave-particle duality, this spin analysis can modify the path information, yielding a recovery of the interference fringes: this is an example of the so-called quantum eraser [31,34]. We define here the quantum eraser simply as the phenomenon where the wave property, i.e. interference, can be reestablished by modifying the path information in the interfering beam after recombining the beams. This quantum eraser was observed in our experiment, for instance, in the case of $\theta=90°$. Interferograms obtained by the O- and the H-beam are plotted for the spinor rotation angles $\alpha=67°$, $90°$, and $180°$ in Fig.7. Here, the path information was labeled in the neutron's spin degree of freedom. Thus, the visibilities of the H-beam (without spin analysis) were reduced according to the degree of this labeling. In particular, no interference oscillation was exhibited in the case of a complete labeling by the orthogonal spinors ($\alpha=90°$). On the contrary, oscillations with visibilities as high as the original (without the spinor rotation) were obtained for the O-beam by completely unlabeling the path information with the use of an appropriate spin analysis. This complete modification of the path information, i.e. complete quantum eraser, can also be interpreted as the post-selection of the sub-ensembles that exhibit perfectly visible fringes, from the ensembles with reduced visibility.

### C. Alternative geometrical description

The experiment presented in this paper is based on two separate evolutions of single quantum state. The choice (6) of different evolutions A and B for the state along the two paths is a technical device to measure the inner product (3) yielding the off-diagonal geometric phase of a fictitious system of two spins which evolve under the action of a single unitary dynamics U. Of course, the same measured phase could also be interpreted directly in terms of geometric property of the actual system under observation. In this section we discuss briefly this alternative interpretation on the Poincaré sphere.

The spectrometer is set up to measure the interference between the two evolved kets:

$$|\Psi_I\rangle = A |\Psi^+\rangle = P(|\Psi^-\rangle) U^{-1} |\Psi^+\rangle, \qquad (13)$$

$$|\Psi_{II}\rangle = B |\Psi^+\rangle = P(|\Psi^-\rangle) U |\Psi^+\rangle. \qquad (14)$$

The first (unitary) part of these evolutions is the $\pm\alpha$ precession, represented by the paths $\Gamma_I$ and $\Gamma_{II}$ on the Poincaré sphere of Fig. 8. The subsequent projections $P(|\Psi^-\rangle)$ are represented by the geodesic lines $G_I$ and $G_{II}$. The standard Berry-Pancharatnam geometric phase is usually defined in terms of a single state, which is compared to itself at the beginning of the evolution. That is realized in practice by taking either A or B to be the identity. We consider the straightforward generalization of letting both A and B be the nontrivial operators of (13) and (14). The generalized Berry-Pancharatnam geometric phase acquired in this direct (two-states) evolution is therefore:

$$\exp(i\gamma^{BP}) = \Phi(\langle\Psi_I|\Psi_{II}\rangle) = \Phi(\{\langle\Psi^+|U\}|\Psi^-\rangle\langle\Psi^-|\{U|\Psi^+\rangle\}), \quad (15)$$

where we have put curly brackets in order to indicate the states on which the unitary evolution is originally acting, and have used the fact that $(U^{-1})^\dagger = U$. This geometric phase is represented by the shaded area surrounded by the two paths starting off at $|\Psi^+\rangle$ and re-joining at $|\Psi^-\rangle$ in Fig. 8 (passing through $|\Psi_I\rangle$ and $|\Psi_{II}\rangle$ respectively). In our experimental circumstances, the solid angle bounded by this loop results in $\Omega = 2\pi + 2\alpha\cdot\cos\theta$, which translates into a geometric phase $\gamma^{BP} = (\pi + \alpha\cdot\cos\theta)$.

These relations are applicable under the condition that U realizes parallel evolution. However, in our experiment (for $\theta \neq \pi/2$) this is not the case: extra dynamical phases contribute as follows:

$$\exp[i(\gamma^{BP} + \Phi_D^I - \Phi_D^{II})] = \Phi(\{\langle\Psi^+|U\}|\Psi^-\rangle\langle\Psi^-|\{U|\Psi^+\rangle\}). \quad (16)$$

By replacing the values of the dynamical phases $\Phi_D^{II/I} = \pm \alpha/2 \cos\theta$, we obtain the observed phase difference

$$\gamma^{BP} + \Phi_D^I - \Phi_D^{II} = (\pi + \alpha\cdot\cos\theta) - \alpha/2 \cos\theta - (-\alpha/2\cos\theta) = \pi, \quad (17)$$

which is indeed the value expected for $\gamma_{+-}^{off}$.

It is worth noting that the dynamical phase difference for the two evolving states does *not* disappear in this "direct-evolution" interpretation of our experiment. On the contrary, the observed phase $\gamma_{+-}^{off}$ is interpreted as the combination of a geometrical part $\gamma^{BP}$ plus a dynamical part $\Phi_D^I - \Phi_D^{II}$. This shades light on the key mechanism on which our single-interference experiment is based. We remind the reader that the definition of the off-diagonal phase would imply the simultaneous evolution of a *pair* of orthogonal states under the action of the same Hamiltonian: that would in principle require a double-interference experiment [11]. The carefully tailored compensation of the dynamical phases of our experiment eliminates one of the interference conditions, providing a way to measure the off-diagonal phase in terms of the standard interference phase of a spin-1/2 system.

### D. Generalizations

The formalism outlined in Sect. II suggests that, in addition to the measurement of "canonical" geometric phases it should be possible to extend the interferometric technique presented in this work to measure other geometric properties of neutron spin evolutions. The choice (6) is indeed a very special one, strictly constraining the motion of the two spinors to remain orthogonal at all times. This is indeed required to measure the geometric properties of the evolution of the whole set of basis states (here $|\Psi^+(t)\rangle$ and $|\Psi^-(t)\rangle$) according to a single dynamics operator U.

However, one could rightfully conceive to let one spinor path evolve under the action of some dynamical operator (for example precession around a $\hat{z}$–oriented magnetic field), while the other spinor evolves according to a completely different dynamics (an $\hat{x}$–oriented magnetic field, say). More complex paths on the Poincaré sphere could be investigated by sequences of differently-oriented magnetic fields. It would be of essential importance to guarantee parallel transport of the states, or a way to explicitly compensate for spurious dynamical phase contributions. If this could be realized, it would be possible to "decouple" the paths of the two spinors on the Poincaré sphere, and thus to investigate a much broader space of possible "relative phases" of the two evolving vectors. In this case, however, the effective space of states visited describes the coherent evolution of two spins at two *different* locations (the two interferometer's paths). Its dimensionality is four, as the space is spanned by the 4 states labeled by spin up/down in the first region, and spin up/down in the second region. Accordingly, an experiment of the kind outlined here can access a more complicated set of diagonal and off-diagonal phases, since a four-state system has a richer range of degrees of freedom then the spin-1/2 system. Obtaining a satisfactory control over these phases could have potential applications in the field of quantum information.

## V. CONCLUSION

In summary, we have exploited spin-polarized neutron interferometry to observe the off-diagonal geometric phase. The essential spinor evolutions were realized by the use of two spin rotators, keeping the directions of the magnetic fields anti-parallel, and the polarization analysis was made after the interferometer. We built a setup insensitive to all (non-geometric) dynamical phases. Final counts with and without the polarization analysis enabled us to directly observe both off-diagonal and diagonal geometric phases in two detectors for noncyclic evolutions. In particular, this experiment provides a direct quantitative measure of the off-diagonal geometric phase. The data confirm the theoretical prediction that this off-diagonal phase should remain constantly equal to $\pi$ in a two-state system. The combined use of the outgoing spin analysis together with the incoming polarized beam extended the range of the experimental conditions: our experimental setup permitted us to measure a wide range of phenomena, including the $4\pi$-periodicity of the spin-1/2 wave function and noncyclic Pancharatnam phase. Furthermore, the results illustrate the significance of the off-diagonal geometric phase, especially when the diagonal geometric phase is undefined. We discuss our experiment in terms of the quantum erasing phenomenon, where the wave property is reestablished by the modification of the path information after recombining the beam in the interferometer. We propose further directions for the investigation of geometric phases in neutron interferometry.

## ACKNOWLEDGEMENTS

We thank R. Gähler for providing needed equipment. This work has been supported by the Austrian Fonds zur Föderung der

# Figure Captions

Fig.1  Schematic representation of the spinor evolution on the Poincaré sphere to illustrate the off-diagonal geometric phase. $|\Psi_i^+\rangle$, evolves to $|\Psi_f^+\rangle$ with corresponding the orthogonal states $|\Psi_i^-\rangle$ to $|\Psi_f^-\rangle$. All states and geodesics are in one plane. The shaded surface corresponds to the off-diagonal geometric phase. Its $2\pi$ solid angle yields a phase shift of $\pi$.

Fig.2  Schematic view of the experimental setup to observe the off-diagonal geometric phase. An incident polarized neutron beam splits into two beam paths in the interferometer and each spinor is rotated to induce appropriate spinor evolutions. *A posteriori* spin analysis allows the observation of the off-diagonal geometric phase. A measurement without spin analysis reveals the diagonal geometric phase.

Fig.3  Typical interferograms for the polar angle of the incident polarization, $\theta=90°$, with various spin rotation angles, $\alpha$. The center figure shows the original ($\alpha=0°$). Curves are the least squares fits. (a) Obtained by the O-beam with a spin analysis. All patterns get the phase shift of 180° with respect to the original when the spinor is rotated, resulting from the off-diagonal geometric phase. (b) Obtained by the H-beam. The patterns show phase shifts of 0° or 180°. These shifts result from a noncyclic Pancharatnam (diagonal) phase.

Fig.4  A series of interferograms for $\theta=60°$ with various $\alpha$, accompanied by curves of the least squares fits. (a) Obtained by the O-beam with a spin analysis. The patterns get the phase shift of 180° to the original when the spinor is rotated, resulting from the off-diagonal geometric phase. (b) Obtained by the H-beam. The patterns get shifted variously, *partly* due to the diagonal geometric phase, or rather due to the noncyclic Pancharatnam phase.

Fig.5 Observed off-diagonal geometric phase for various polar angle θ, of the incident polarization. In case of ideal experimental situation, theory predicts the off-diagonal geometric phase of 180° except for the point α=0°, where it is undefined. Dashed lines are the theoretical predictions with effective incident polarizations taken into account.

Fig.6 Intensity modulations obtained by the O- and the H-beams to illustrate the significance of the off-diagonal geometric phase, for 180° spinor rotation. While the H-beam intensity shows no interference oscillations due to the undefined diagonal geometric phase, the O-beam intensity shows a remarkable oscillation since the off-diagonal geometric phases has a definite value.

Fig.7 Intensity modulations obtained by the O- and the H-beams to illustrate a quantum eraser for the polar angle of the incident spinor θ=90° and spinor rotation angles (a) α=67°, (b) 90°, and (c) 180°. When the path information is labeled in the neutron's spin degree of freedom, the visibilities of the oscillations are reduced (H-beam). In contrast, when this labeling is erased with the use of the spin analysis, oscillations with high visibilities come out (O-beam).

Fig.8 Schematic representation of the spinor evolution on the spin sphere to illustrate alternative interpretation of the off-diagonal geometric phase: a combination of the geometric and dynamical parts. at $|\Psi^+\rangle$ evolves to $|\Psi_I\rangle$ along the trajectory $\Gamma_I$ and to $|\Psi_{II}\rangle$ along $\Gamma_{II}$ and ends up to $|\Psi^-\rangle$ along the geodesic paths $G_I$ and $G_{II}$. The shaded surface corresponds to the geometric part of the off-diagonal geometric phase.

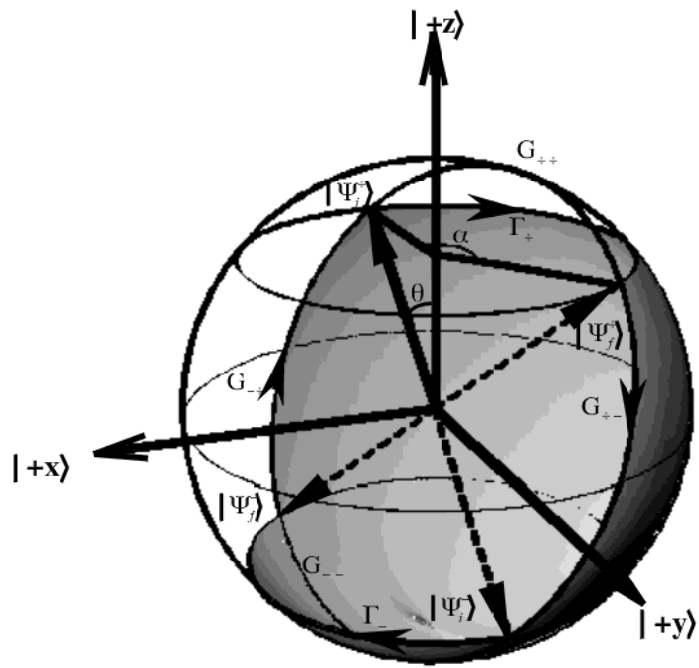

Fig.1 Hasegawa et al.

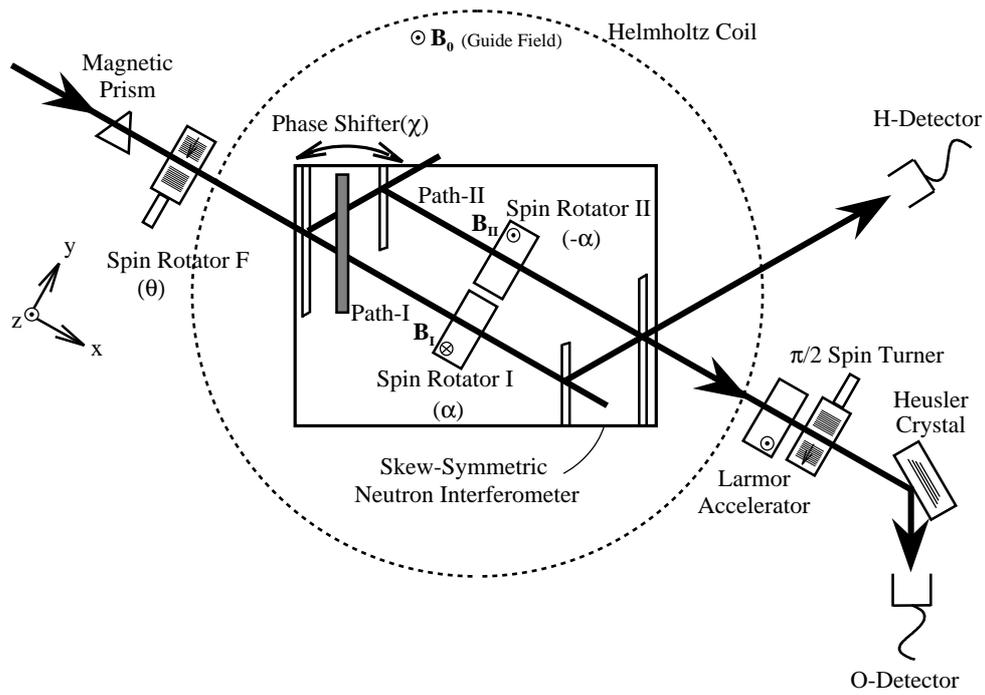

Fig.2   Hasegawa et al.

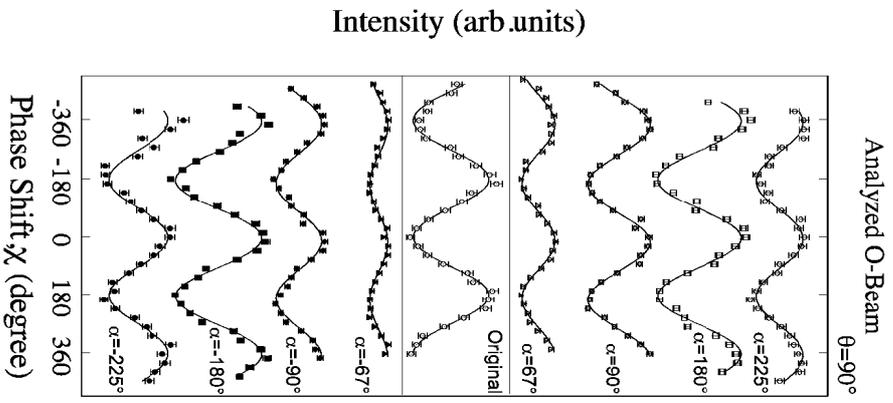
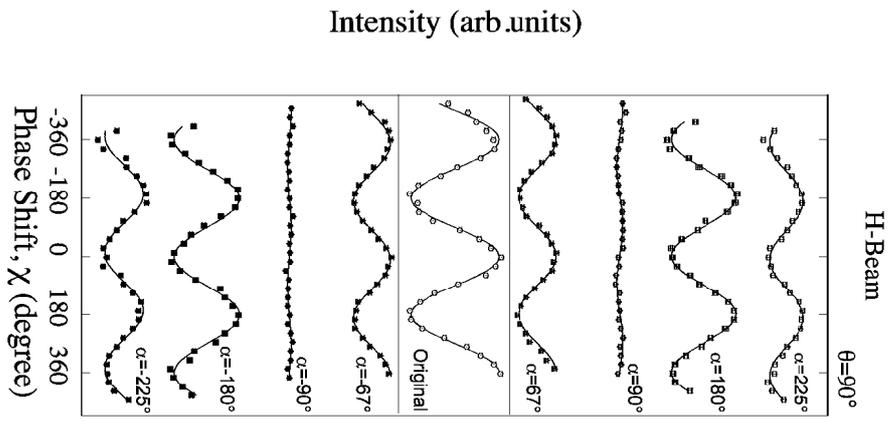

Fig.3 Hasegawa et al.

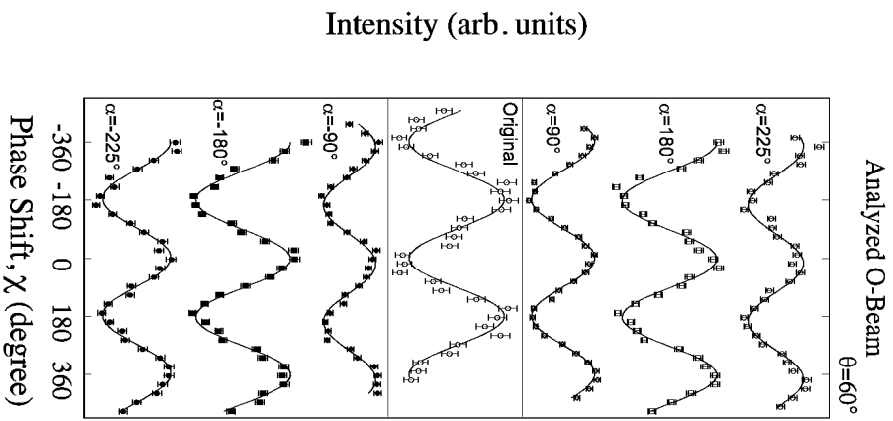

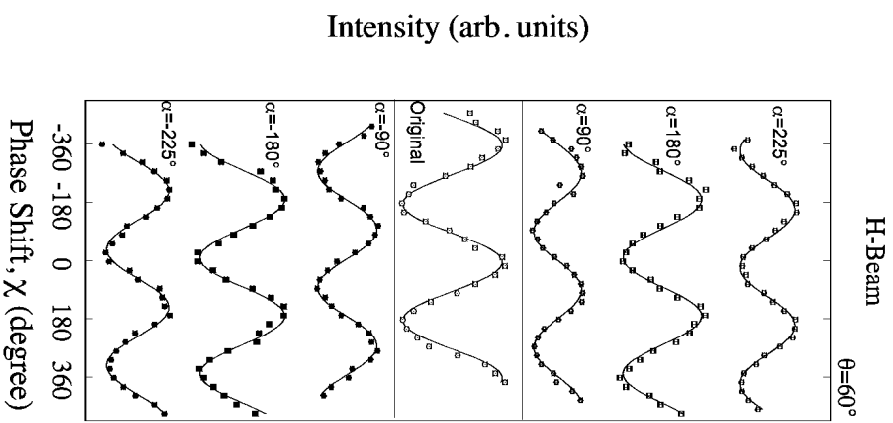

Fig.4 Hasegawa et al.

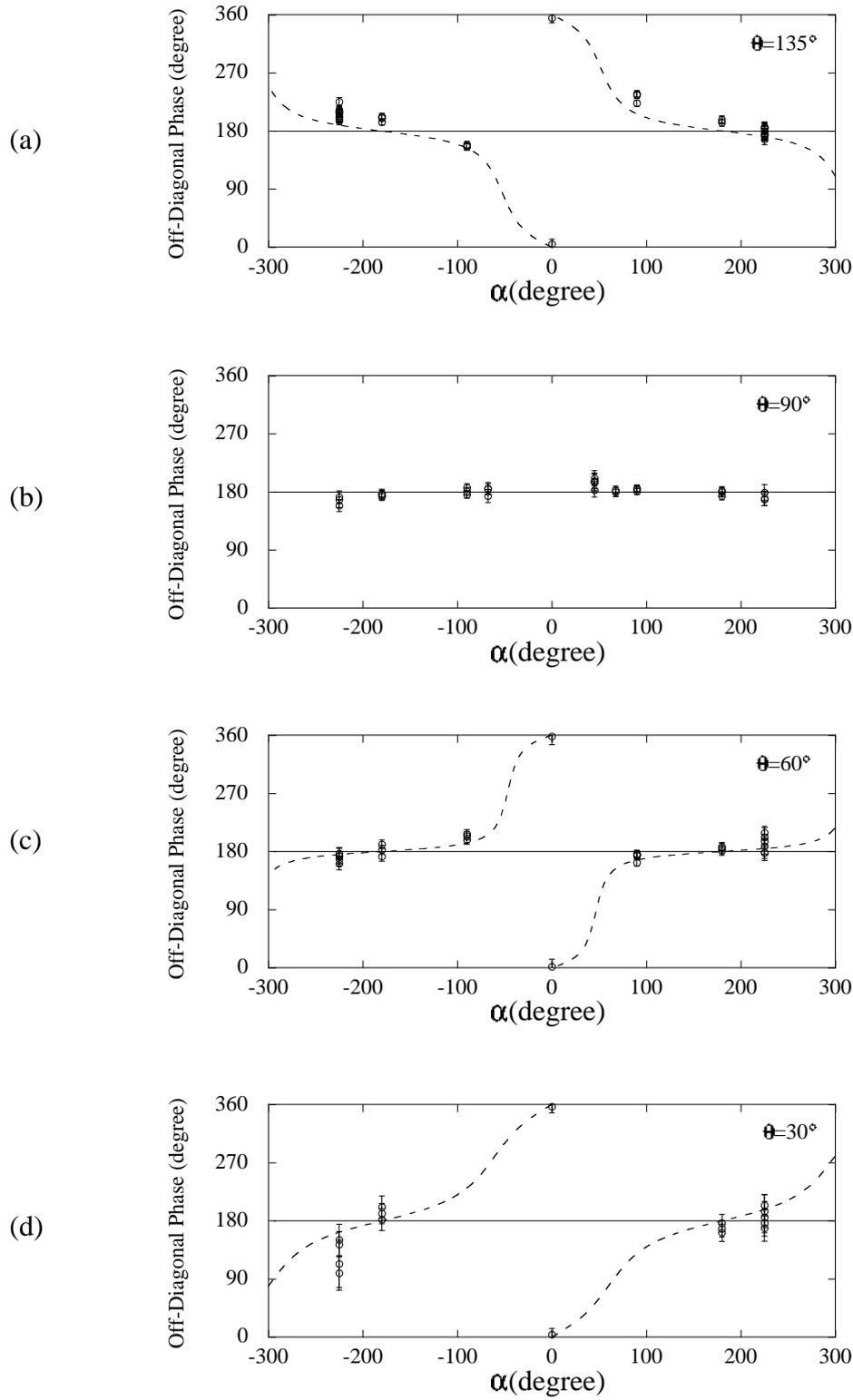

Fig.5 Hasegawa et al.

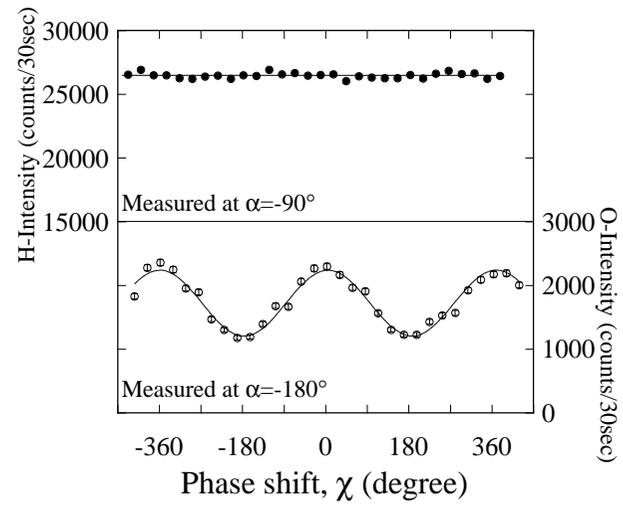

Fig.6 Hasegawa et al.

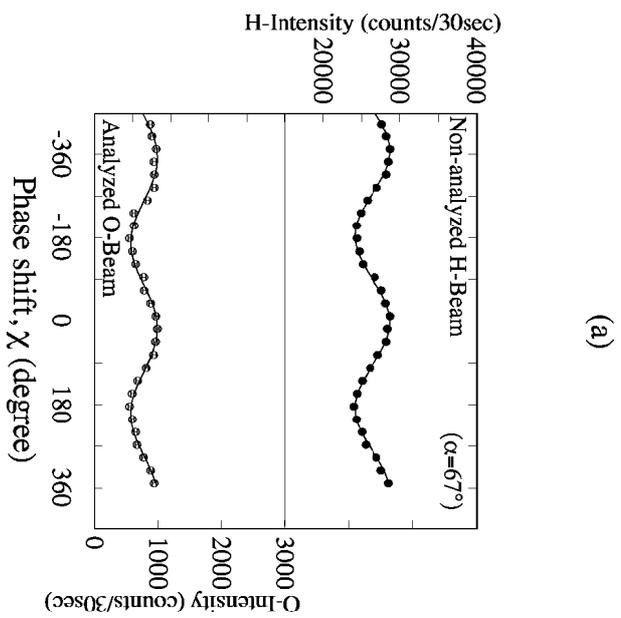

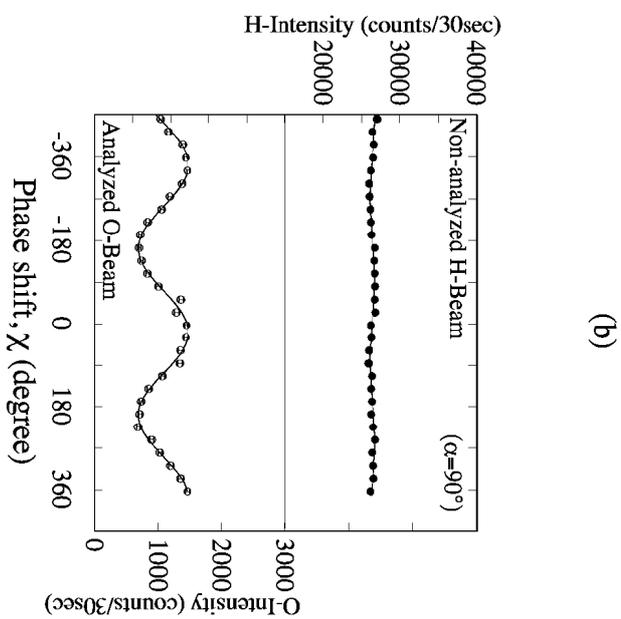

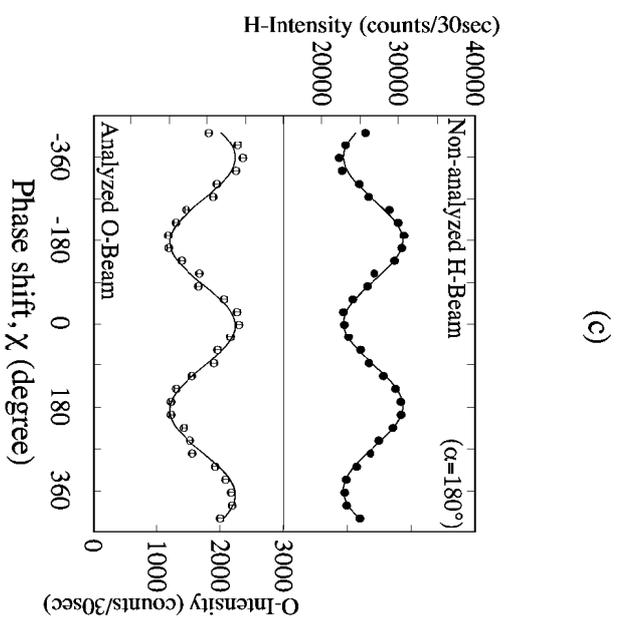

Fig.7 Hasegawa et al.

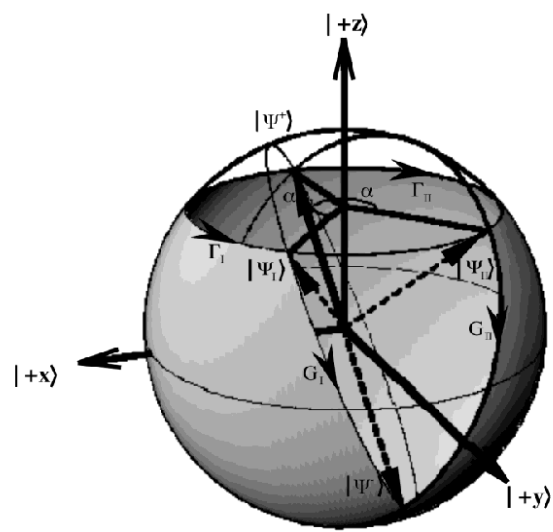

Fig.8 Hasegawa et al.